\documentclass[11pt]{article}

\usepackage[dvipsnames]{xcolor}
\usepackage{framed}
\definecolor{shadecolor}{rgb}{0.9,0.9,0.9}

\usepackage{mathtools}
\usepackage{amsmath}
\usepackage[shortlabels]{enumitem}

\usepackage{graphicx,epic,eepic,epsfig,amsmath,latexsym,amssymb,verbatim,color}
 
\usepackage{amsfonts}       
\usepackage{nicefrac}       

\usepackage{amsmath}
\usepackage{bbm}

\usepackage{float}
\usepackage{tikz}
\usetikzlibrary{chains}
\usetikzlibrary{fit}
\usepackage{pgflibraryarrows}		
\usepackage{pgflibrarysnakes}		

\usepackage{epsfig}
\usetikzlibrary{shapes.symbols,patterns} 
\usepackage{pgfplots}

\usepackage[strict]{changepage}
\usepackage{hyperref}
\hypersetup{colorlinks=true,citecolor=blue,linkcolor=blue,filecolor=blue,urlcolor=blue,breaklinks=true}

\usepackage[marginal]{footmisc}
\usepackage{url}
\usepackage{theorem}
 
\def\squareforqed{\hbox{\rlap{$\sqcap$}$\sqcup$}}
\def\qed{\ifmmode\squareforqed\else{\unskip\nobreak\hfil
\penalty50\hskip1em\null\nobreak\hfil\squareforqed
\parfillskip=0pt\finalhyphendemerits=0\endgraf}\fi}
\def\endenv{\ifmmode\;\else{\unskip\nobreak\hfil
\penalty50\hskip1em\null\nobreak\hfil\;
\parfillskip=0pt\finalhyphendemerits=0\endgraf}\fi}

\newcounter{remark}

\newcounter{example}

\mathchardef\ordinarycolon\mathcode`\:
\mathcode`\:=\string"8000
\def\vcentcolon{\mathrel{\mathop\ordinarycolon}}
\begingroup \catcode`\:=\active
  \lowercase{\endgroup
  \let :\vcentcolon
  }

\usepackage{cleveref}
\usepackage{graphicx}
\usepackage{xcolor}

\RequirePackage[framemethod=default]{mdframed}
\newmdenv[skipabove=7pt,
skipbelow=7pt,
backgroundcolor=darkblue!15,
innerleftmargin=5pt,
innerrightmargin=5pt,
innertopmargin=5pt,
leftmargin=0cm,
rightmargin=0cm,
innerbottommargin=5pt,
linewidth=1pt]{tBox}

\newmdenv[skipabove=7pt,
skipbelow=7pt,
backgroundcolor=red!15,
innerleftmargin=5pt,
innerrightmargin=5pt,
innertopmargin=5pt,
leftmargin=0cm,
rightmargin=0cm,
innerbottommargin=5pt,
linewidth=1pt]{rBox}

\newmdenv[skipabove=7pt,
skipbelow=7pt,
backgroundcolor=blue2!25,
innerleftmargin=5pt,
innerrightmargin=5pt,
innertopmargin=5pt,
leftmargin=0cm,
rightmargin=0cm,
innerbottommargin=5pt,
linewidth=1pt]{dBox}
\newmdenv[skipabove=7pt,
skipbelow=7pt,
backgroundcolor=darkkblue!15,
innerleftmargin=5pt,
innerrightmargin=5pt,
innertopmargin=5pt,
leftmargin=0cm,
rightmargin=0cm,
innerbottommargin=5pt,
linewidth=1pt]{sBox}
\definecolor{darkblue}{RGB}{0,76,156}
\definecolor{darkkblue}{RGB}{0,0,153}
\definecolor{blue2}{RGB}{102,178,255}
\definecolor{darkred}{RGB}{195,0,0}

\newcommand{\nc}{\newcommand}
\nc{\rnc}{\renewcommand}
\nc{\lbar}[1]{\overline{#1}}
\nc{\bra}[1]{\langle#1|}
\nc{\ket}[1]{|#1\rangle}
\nc{\ketbra}[2]{|#1\rangle\!\langle#2|}
\nc{\braket}[2]{\langle#1|#2\rangle}

\nc{\proj}[1]{| #1\rangle\!\langle #1 |}
\nc{\avg}[1]{\langle#1\rangle}
\nc{\smfrac}[2]{\mbox{$\frac{#1}{#2}$}}
\nc{\tr}{\operatorname{Tr}}
\nc{\ox}{\otimes}
\nc{\dg}{\dagger}
\nc{\dn}{\downarrow}
\nc{\cA}{{\cal A}}
\nc{\cB}{{\cal B}}
\nc{\cC}{{\cal C}}
\nc{\cD}{{\cal D}}
\nc{\cE}{{\cal E}}
\nc{\cF}{{\cal F}}
\nc{\cG}{{\cal G}}
\nc{\cH}{{\cal H}}
\nc{\cI}{{\cal I}}
\nc{\cJ}{{\cal J}}
\nc{\cK}{{\cal K}}
\nc{\cL}{{\cal L}}
\nc{\cM}{{\cal M}}
\nc{\cN}{{\cal N}}
\nc{\cO}{{\cal O}}
\nc{\cP}{{\cal P}}
\nc{\cQ}{{\cal Q}}
\nc{\cR}{{\cal R}}
\nc{\cS}{{\cal S}}
\nc{\cT}{{\cal T}}
\nc{\cU}{{\cal U}}
\nc{\cV}{{\cal V}}
\nc{\cX}{{\cal X}}
\nc{\cY}{{\cal Y}}
\nc{\cZ}{{\cal Z}}
\nc{\cW}{{\cal W}}
\nc{\csupp}{{\operatorname{csupp}}}
\nc{\qsupp}{{\operatorname{qsupp}}}
\nc{\var}{{\operatorname{var}}}
\nc{\rar}{\rightarrow}
\nc{\lrar}{\longrightarrow}
\nc{\polylog}{{\operatorname{polylog}}}
\nc{\wt}{{\operatorname{wt}}}
\nc{\av}[1]{{\left\langle {#1} \right\rangle}}
\nc{\supp}{{\operatorname{supp}}}

\nc{\argmin}{{\operatorname{argmin}}}

\def\x{\xi}

\nc{\RR}{{{\mathbb R}}}
\nc{\CC}{{{\mathbb C}}}
\nc{\FF}{{{\mathbb F}}}
\nc{\NN}{{{\mathbb N}}}
\nc{\ZZ}{{{\mathbb Z}}}
\nc{\PP}{{{\mathbb P}}}
\nc{\QQ}{{{\mathbb Q}}}
\nc{\UU}{{{\mathbb U}}}
\nc{\EE}{{{\mathbb E}}}
\nc{\id}{{\operatorname{id}}}

\nc{\CHSH}{{\operatorname{CHSH}}}

\nc{\be}{\begin{equation}}
\nc{\ee}{{\end{equation}}}
\nc{\bea}{\begin{eqnarray}}
\nc{\eea}{\end{eqnarray}}
\nc{\<}{\langle}
\rnc{\>}{\rangle}
\nc{\rU}{\mbox{U}}

\nc{\ob}[1]{#1}

\nc{\SEP}{{\text{\rm SEP}}}
\nc{\NS}{{\text{\rm NS}}}
\nc{\LOCC}{{\text{\rm LOCC}}}
\nc{\PPT}{{\text{\rm PPT}}}
\nc{\EXT}{{\text{\rm EXT}}}
\nc{\Sym}{{\operatorname{Sym}}}

\nc{\ERLO}{{E_{\text{r,LO}}}}
\nc{\ERLOCC}{{E_{\text{r,LOCC}}}}
\nc{\ERPPT}{{E_{\text{r,PPT}}}}
\nc{\ERLOCCinfty}{{E^{\infty}_{\text{r,LOCC}}}}
\nc{\Aram}{{\operatorname{\sf A}}}

\usepackage{tikz}
\usepackage{hyperref}
\hypersetup{colorlinks=true,citecolor=blue,linkcolor=blue,filecolor=blue,urlcolor=blue,breaklinks=true}

\makeatletter
\def\grd@save@target#1{%
  \def\grd@target{#1}}
\def\grd@save@start#1{%
  \def\grd@start{#1}}
\tikzset{
  grid with coordinates/.style={
    to path={%
      \pgfextra{%
        \edef\grd@@target{(\tikztotarget)}%
        \tikz@scan@one@point\grd@save@target\grd@@target\relax
        \edef\grd@@start{(\tikztostart)}%
        \tikz@scan@one@point\grd@save@start\grd@@start\relax
        \draw[minor help lines,magenta] (\tikztostart) grid (\tikztotarget);
        \draw[major help lines] (\tikztostart) grid (\tikztotarget);
        \grd@start
        \pgfmathsetmacro{\grd@xa}{\the\pgf@x/1cm}
        \pgfmathsetmacro{\grd@ya}{\the\pgf@y/1cm}
        \grd@target
        \pgfmathsetmacro{\grd@xb}{\the\pgf@x/1cm}
        \pgfmathsetmacro{\grd@yb}{\the\pgf@y/1cm}
        \pgfmathsetmacro{\grd@xc}{\grd@xa + \pgfkeysvalueof{/tikz/grid with coordinates/major step}}
        \pgfmathsetmacro{\grd@yc}{\grd@ya + \pgfkeysvalueof{/tikz/grid with coordinates/major step}}
        \foreach \x in {\grd@xa,\grd@xc,...,\grd@xb}
        \node[anchor=north] at (\x,\grd@ya) {\pgfmathprintnumber{\x}};
        \foreach \y in {\grd@ya,\grd@yc,...,\grd@yb}
        \node[anchor=east] at (\grd@xa,\y) {\pgfmathprintnumber{\y}};
      }
    }
  },
  minor help lines/.style={
    help lines,
    step=\pgfkeysvalueof{/tikz/grid with coordinates/minor step}
  },
  major help lines/.style={
    help lines,
    line width=\pgfkeysvalueof{/tikz/grid with coordinates/major line width},
    step=\pgfkeysvalueof{/tikz/grid with coordinates/major step}
  },
  grid with coordinates/.cd,
  minor step/.initial=.2,
  major step/.initial=1,
  major line width/.initial=2pt,
}
\makeatother

\usepackage{thmtools}
\usepackage{thm-restate}
\usepackage{etoolbox}
\makeatletter
\def\problem@s{}
\newcounter{problems@cnt}

\newcommand{\allproblems}{\problem@s}
\makeatother

\newcommand{\Place}[3]{
\begin{scope}[xshift = #2cm, yshift = #3cm]
  #1
\end{scope}}
\newcommand{\place}[3]{\Place{\csuse{#1}}{#2}{#3}}

\newcommand{\round}{.. controls +(0.3,0) and +(-0.3,0) ..}

\csdef{S1}{
  \draw (0, \Y)   --   (\X, \Y);
  \draw (0,  0)   --   (\X,  0);
  \draw (0,-\Y)   --   (\X,-\Y);
}
\csdef{S2}{
  \draw (0, \Y) \round (\X,  0);
  \draw (0,  0) \round (\X,-\Y);
  \draw (0,-\Y) \round (\X, \Y);
}
\csdef{S3}{
  \draw (0, \Y) \round (\X,-\Y);
  \draw (0,  0) \round (\X, \Y);
  \draw (0,-\Y) \round (\X,  0);
}
\csdef{S4}{
  \draw (0, \Y)   --   (\X, \Y);
  \draw (0,  0) \round (\X,-\Y);
  \draw (0,-\Y) \round (\X,  0);
}
\csdef{S5}{
  \draw (0, \Y) \round (\X,  0);
  \draw (0,  0) \round (\X, \Y);
  \draw (0,-\Y)   --   (\X,-\Y);
}
\csdef{S6}{
  \draw (0, \Y) \round (\X,-\Y);
  \draw (0,  0)   --   (\X,  0);
  \draw (0,-\Y) \round (\X, \Y);
}

\csdef{T1}{\csuse{S1}}
\csdef{T2}{\csuse{S3}} 
\csdef{T3}{\csuse{S2}} 
\csdef{T4}{\csuse{S4}}
\csdef{T5}{\csuse{S5}}
\csdef{T6}{\csuse{S6}}

\csdef{Bx}#1#2#3{
  \place{S1}{1*\X}{0}
  \node[box] at (1.5*\X,-\Y) {\scriptsize };
}
\csdef{Bxs}{\Bx{1}{2}{3}}

\csdef{Tri}#1#2#3{
  \place{S1}{1*\X}{0}
  \node[tri] at (1.5*\X, \Y) {\scriptsize #1};
  \node[tri] at (1.5*\X,  0) {\scriptsize #2};
  \node[tri] at (1.5*\X,-\Y) {\scriptsize #3};
}

\newcommand{\IOO}[3]{
  \draw (-0.2*\X,0) -- (3.2*\X,0);
  \node[circle] at (1.5*\X,  0) {\scriptsize  };
  \node[circle] at (0.5*\X,-\Y) {\scriptsize };
  \node at (1.5*\X,-\Y) {\scriptsize $d\delta_{ij}$};
}
\csdef{C11}{\IOO{1}{2}{3}}
\csdef{C22}{\IOO{2}{1}{3}}
\csdef{C44}{\IOO{1}{2}{3}}
\csdef{C55}{\IOO{1}{2}{3}}

\newcommand{\IOOA}[3]{
  \draw (-0.2*\X,0) -- (3.2*\X,0);
  \node[box, fill=red!30] at (1.5*\X,  0) {\scriptsize  };
  \node[circle] at (0.5*\X,-\Y) {\scriptsize };
  \node[circle] at (1.5*\X,-\Y) {\scriptsize $d^2$ };

}
\csdef{C33}{\IOOA{1}{2}{3}}
\csdef{C66}{\IOOA{1}{2}{3}}

\newcommand{\IIO}[3]{
  \draw (-0.2*\X,0) -- (3.2*\X,0);
  \node[box, fill=red!30] at (  1.5*\X,  0) {\scriptsize };
 \node[circle] at (1.5*\X,-\Y) {\scriptsize $d$};
  \node[circle] at (2.5*\X,-\Y) {\scriptsize };
}

\csdef{C61}{\IIO{2}{1}{3}}
\csdef{C43}{\IIO{2}{1}{3}}
\csdef{C53}{\IIO{2}{1}{3}}
\csdef{C62}{\IIO{2}{1}{3}}

\newcommand{\IIOA}[3]{
  \draw (-0.2*\X,0) -- (3.2*\X,0);
  \node[circle] at (  1.5*\X,  0) {\scriptsize };
 \node[circle] at (0.5*\X,-\Y) {\scriptsize };
  \node[circle] at (1.5*\X,-\Y) {\scriptsize $\delta_{ij}$};
}
\csdef{C51}{\IIOA{1}{2}{3}}
\csdef{C42}{\IIOA{1}{2}{3}}

\newcommand{\OOO}[3]{
  \draw (-0.2*\X,0) -- (3.2*\X,0);
  \node[circle] at (1.5*\X,  0) {\scriptsize };
  \node[circle] at (  1*\X,-\Y) {\scriptsize };
  \node[circle] at (  1.5*\X,-\Y) {\scriptsize $\delta_{ij}$};
}
\csdef{C41}{\OOO{1}{2}{3}}
\csdef{C52}{\OOO{1}{2}{3}}

\newcommand{\OOOA}[3]{
  \draw (-0.2*\X,0) -- (3.2*\X,0);
  \node[box, fill=red!30] at (1.5*\X,  0) {\scriptsize };
  \node[circle] at (  1*\X,-\Y) {\scriptsize };
  \node[circle] at (  1.5*\X,-\Y) {\scriptsize $d$};
}
\csdef{C63}{\OOOA{1}{2}{3}}

\newcommand{\III}[3]{
  \draw (-0.2*\X,-0.5*\Y) -- (3.2*\X,-0.5*\Y);
  \node[box, fill=red!30] at (1.5*\X,-0.5*\Y) {\scriptsize };
}
\csdef{C21}{\III{2}{3}{1}}
\csdef{C31}{\III{3}{2}{1}}
\csdef{C32}{\III{3}{1}{2}}
\csdef{C54}{\III{2}{3}{1}}
\csdef{C64}{\III{3}{2}{1}}
\csdef{C65}{\III{3}{1}{2}}

\usepackage{amsmath,amsfonts,amssymb,array,graphicx,mathtools,multirow,bm,times,tcolorbox,relsize,booktabs}
\usepackage[utf8]{inputenc}
\usepackage[T1]{fontenc}
\usepackage{qcircuit}
\usepackage{authblk} 
\usepackage[paperwidth=199.8mm,paperheight=297mm,centering,hmargin=20mm,vmargin=2cm]{geometry}
\usepackage[ruled, vlined, linesnumbered]{algorithm2e}
\usepackage{tikz}
\usetikzlibrary{arrows.meta,positioning,shapes,calc}

\definecolor{colortwo}{rgb}{0.4,0.77,0.17}
\definecolor{colorthree}{rgb}{0.01,0.51,0.93}

\allowdisplaybreaks

\begin{document}
\title{Block Coordinate Descent for Dynamic Portfolio Optimization on Finite-Precision Coherent Ising Machines}

\author[1]{Keming He \thanks{These authors contributed equally to this work.}}
\author[1,2]{Yuehan Zhang \protect\footnotemark[1]}
\author[1]{Hongshun Yao\thanks{yaohongshun2021@gmail.com}}
\author[3]{Jin-Guo Liu\thanks{jinguoliu@hkust-gz.edu.cn}}
\author[1]{Xin Wang
\thanks{felixxinwang@hkust-gz.edu.cn}}

\affil[1]{\small Thrust of Artificial Intelligence, Information Hub,\par The Hong Kong University of Science and Technology (Guangzhou), Guangdong 511453, China.}
\affil[2]{\small School of Cyber Science and Engineering, Sichuan University, Sichuan 610065, China.}
\affil[3]{\small Thrust of Advanced Materials, Information Hub,\par The Hong Kong University of Science and Technology (Guangzhou), Guangdong 511453, China.}

\date{\today}
\maketitle

\begin{abstract}


Coherent Ising machines~(CIMs) have emerged as specialized quantum hardware for large-scale combinatorial optimization. However, for large instances that remain challenging for classical methods, some platforms support only finite-precision inputs, and the required scaling and quantization can degrade solution quality. Dynamic portfolio optimization~(DPO) can be formulated as a quadratic unconstrained binary optimization~(QUBO) problem, but large instances are especially vulnerable to precision loss under global scaling. We propose a block coordinate descent method that decomposes the DPO model along the time dimension and iteratively solves compact time-block subproblems on the device. Experiments on finite-precision CIM hardware show that the method enables these instances to be solved under hardware precision limits, yields portfolios competitive with classical benchmark solvers, and reduces runtime through fast CIM solution of the resulting subproblems. These results demonstrate the promise of finite-precision CIMs as a practical and scalable approach to structured large-scale combinatorial optimization.

\end{abstract}






\section{Introduction}

Quantum computing processors have achieved experimental demonstrations of beyond classical performance on specialized tasks, showing that quantum hardware can outperform classical computation in selected regimes \cite{arute2019quantum,zhong2020quantum,king2025beyond,ai2025observation, ebadi2022quantum}. These milestones have accelerated efforts to identify application classes where such advantages can translate into practical computational gains beyond proof of principle demonstrations. Combinatorial optimization is a leading target because a wide range of discrete decision problems admit polynomial reductions to minimizing an Ising energy or equivalently to a QUBO formulation \cite{lucas2014ising}. This common formulation provides a shared interface between applications and optimization backends. It also supports a broad set of quantum and quantum-inspired approaches, including quantum annealing \cite{johnson2011quantum} and gate-based variational algorithms such as the quantum approximate optimization algorithm and the variational quantum eigensolver \cite{farhi2014quantum,peruzzo2014variational}.

Dynamic portfolio optimization~(DPO) seeks an optimal sequence of trades over multiple rebalancing times to maximize return while controlling risk and accounting for transaction costs and other constraints \cite{mugel2022dynamic,Nodar2024ScalingTV}. Inter temporal couplings induced by rebalancing and trading frictions make the problem substantially harder than single period variants. With standard binary encodings, the resulting QUBO or Ising instances are typically large and dense, scaling with the number of assets, the number of time steps, and the bit depth used to represent holdings. These characteristics make DPO a natural candidate for quantum and quantum-inspired optimization, and multiple studies have investigated QUBO based workflows using gate-based variational algorithms such as QAOA and VQE \cite{mugel2022dynamic,carrascal2023backtesting,kerenidis2019quantum,buonaiuto2023best, Nodar2024ScalingTV}.

Coherent Ising machines~(CIMs) are photonic Ising optimizers that represent binary spins by the phases of degenerate optical parametric oscillators and implement programmable couplings through measurement and feedback \cite{mcmahon2016fully,inagaki2016coherent}. CIM experiments have demonstrated the capability to solve large, dense Max-Cut problems and have reported substantial speed advantages over classical solvers on the studied benchmarks \cite{inagaki2016coherent,honjo2021100}. Recent work has also explored the use of CIM for practical combinatorial optimization problems~\cite{guo2025disaggregation, ju2025quantum}. These results position CIMs as a promising hardware platform for large-scale combinatorial optimization. However, deploying standard QUBO formulations on some CIM devices is not straightforward, because the hardware accepts only finite, quantized input coefficients over a limited dynamic range. Large QUBO models must therefore be globally rescaled and quantized before execution, which can suppress weaker but important interactions and alter the effective optimization landscape. This issue is particularly acute for DPO, where objective and constraint terms can differ substantially in magnitude, so naive scaling and quantization can materially degrade solution quality.

In this work, we propose a precision-aware block coordinate descent~(BCD) strategy for the DPO model~\cite{tseng2001convergence,beck2013convergence}. BCD decomposes the global DPO QUBO along the time dimension, solves a compact subproblem for one time block on the device while treating the other blocks as fixed conditions, and iterates until a satisfactory solution is reached. By reducing both the effective problem size and the coefficient range presented in each device call, the method alleviates the need for aggressive global scaling. We evaluate the proposed scheme on Qboson CIM hardware~\cite{qboson_kaiwu_sdk_2025}, where direct global INT8 execution is infeasible on all tested instances, while our BCD-based workflow yields feasible solutions with improved solution quality. The resulting portfolios are comparable to classical benchmark methods including simulated annealing and tabu search, while reducing the measured solver runtime by several orders of magnitude on the tested instances, with observed gains over \(10^5\)\(\times\). Overall, the proposed method provides a practical and robust route toward scalable CIM-based dynamic portfolio optimization  under finite-precision constraints.

\section{Background and motivation}

\subsection{Dynamic portfolio optimization}

Portfolio optimization concerns the allocation of capital across multiple assets in order to balance expected return against risk. In the mean--variance framework of Markowitz, risk is quantified by the variance of portfolio returns, and the goal is to identify trade-offs between risk and return. In this work we adopt the dynamic portfolio optimization (DPO) setting of~\cite{Nodar2024ScalingTV, mugel2022dynamic, rosenberg2015solving}.

We consider a portfolio with $N_a$ tradable assets rebalanced over $N_t$ decision times. Let $\omega_{t,a}$ denote the capital invested in asset $a\in\{0,\dots,N_a-1\}$ at rebalancing time $t\in\{0,\dots,N_t-1\}$, and let $\Omega\in\mathbb{R}^{N_t\times N_a}$ denote the allocation matrix with entries $(\Omega)_{t,a}=\omega_{t,a}$. The objective comprises four components: expected return $F(\Omega)$, risk $R(\Omega)$, transaction costs $C(\Omega)$, and budget-constraint penalties $B(\Omega)$. The resulting objective is
\begin{equation}\label{eq:objective}
    O(\Omega) = F(\Omega) - R(\Omega) - C(\Omega) - B(\Omega) .
\end{equation}
We define the four terms below. The expected return is defined as
\begin{equation}
    F(\Omega)=\sum_{t=0}^{N_t-1}\sum_{a=0}^{N_a-1}\omega_{t,a}\,\mu_{t,a},
\end{equation}
where $\mu_{t,a}$ denotes the logarithmic return of asset $a$ over the period $[t,t+1]$. Let $P_{t,a}$ be the price of asset $a$ at time $t$, and we compute
\begin{equation}
\mu_{t,a}=\log\!\left(\frac{P_{t+1,a}}{P_{t,a}}\right).
\end{equation}
Transaction costs arise from rebalancing. We model them as proportional to the amount traded,
\begin{equation}
    \nu \sum_{t=0}^{N_t-1}\sum_{a=0}^{N_a-1}\left|\omega_{t,a}-\omega_{t-1,a}\right|,
\end{equation}
where $\nu>0$ is the transaction-cost rate and we set $\omega_{-1,a}=0$ to represent an initially uninvested portfolio. Since the absolute value is non-quadratic, we approximate it by a quadratic surrogate,
\begin{equation}
    C(\Omega)= \nu\lambda \sum_{t=0}^{N_t-1}\sum_{a=0}^{N_a-1}\left(\omega_{t,a}-\omega_{t-1,a}\right)^2,
\end{equation}
where $\lambda>0$ controls the strength of the approximation.  
We impose a total budget $K$ at each time:
\begin{equation}\label{eq:budget-constraints}
    \sum_{a=0}^{N_a-1}\omega_{t,a}=K,\qquad \forall t.
\end{equation}
The budget constraint can then be enforced via the quadratic penalty
\begin{equation}
    B(\Omega)=\rho \sum_{t=0}^{N_t-1}\left(\sum_{a=0}^{N_a-1}\omega_{t,a}-K\right)^2,
\end{equation}
where $\rho>0$ is a penalty coefficient.  

The estimation of risk should depend on the change of price within a rebalancing time. Since each decision at time $t$ remains in effect until the next rebalance, we estimate risk from daily price fluctuations observed during that interval. Let $\Delta t$ denote the number of daily return observations in a rebalancing interval, and let $P_{s,a}$ be the closing price of asset $a$ on day $s$. The daily logarithmic return is
\begin{equation}
    \mu^{d}_{s,a}=\log\!\left(\frac{P_{s+1,a}}{P_{s,a}}\right).
\end{equation}
For each rebalancing time, define the set of daily indices in that interval $\mathcal{S}_t=\{t\Delta t,\; t\Delta t+1,\; \dots,\; (t+1)\Delta t-1\}.$ We consider three standard covariance-based estimators for the risk matrix at time $t$.

\textit{Covariance.}
The covariance matrix is defined as
\begin{equation}
    \Sigma_t^{\text{cov}}(a,b)=\frac{1}{\Delta t-1}\sum_{s\in\mathcal{S}_t}
    \big(\mu^{d}_{s,a}-\bar{\mu}^{d}_{t,a}\big)\big(\mu^{d}_{s,b}-\bar{\mu}^{d}_{t,b}\big),
\end{equation}
where $ \bar{\mu}^{d}_{t,a}$ is the average daily logarithmic return 
\begin{equation}
    \bar{\mu}^{d}_{t,a}=\frac{1}{\Delta t}\sum_{s\in\mathcal{S}_t}\mu^{d}_{s,a}.
\end{equation}
This is the typical choice in the Markowitz framework. However, it treats upside and downside fluctuations symmetrically. The estimate can be noisy when $\Delta t$ is limited, which may cause allocations to vary sharply across rebalancing times.

\textit{Semicovariance.}
In order to penalize primarily downside moves, one may replace covariance by semicovariance with respect to a benchmark $B$~\cite{Estrada2007MeansemivarianceBD,Estrada2007MeanSemivarianceOA}:
\begin{equation}
    \Sigma^{\text{semi}}_t(a,b )=\frac{1}{\Delta t-1}\sum_{s\in\mathcal{S}_t}
    \min\!\big(\mu^{d}_{s,a}-B,0\big)\cdot \min\!\big(\mu^{d}_{s,b}-B,0\big).
\end{equation}
We choose $B=0$ in our experiments for simplicity. Intuitively, semicovariance focuses attention on periods when returns fall below the benchmark, and ignores the movement during days with positive return.

\textit{Shrinkage covariance.}
To make risk estimation more stable, we can also consider a linear shrinkage estimator~\cite{ledoit2003honey,ledoit2003improved,Ledoit2004AWE}
\begin{equation}
    \Sigma_t^{\text{sh}} (a,b)=(1-\delta)\Sigma_t^{\text{cov}}(a,b)+\delta F_t,
\end{equation}
where $F_t$ is a structured target matrix, and $\delta \in [0,1]$ is the shrinkage intensity. For simplicity, we take $F_t$ proportional to identity matrix,  $F_t=\frac{\mathrm{tr}(\Sigma_t^{\text{cov}})}{N_a} I$, and $\delta=\mathrm{clip}\!\left(\frac{\hat{\beta}}{\hat{\alpha}},\,0,\,1\right)$ to minimize the expected squared estimation error in Frobenius norm, where  $\hat{\alpha}$ measures how far the sample covariance is from the target, and $\hat{\beta}$ estimates the aggregate sampling noise of the sample covariance entries in $\Sigma_t$. 

In what follows we restrict to $\Sigma_t\in\{\Sigma_t^{\mathrm{cov}},\,\Sigma_t^{\mathrm{semi}},\,\Sigma_t^{\mathrm{sh}}\}$. These three choices capture complementary aspects that are most relevant for our setting.
More sophisticated risk models such as conditional value-at-risk~(CVaR)~\cite{rockafellar2000optimization}, conditional drawdown at risk~(CDaR)~\cite{chekhlov2005drawdown}, or conditional-volatility models~\cite{engle1986modelling} are also well motivated, but they typically introduce non-quadratic objectives or additional state variables and hyperparameters, which increases modeling and computational complexity beyond the scope of this work. Given a choice of $\Sigma_t$, the risk term is
\begin{equation}\label{eq:risk_variance}
    R(\Omega)=\frac{\gamma}{2}\sum_{t=0}^{N_t-1}\sum_{a,b=0}^{N_a-1}
\omega_{t,a}\,\Sigma_t(a,b)\,\omega_{t,b},
\end{equation}
where $\gamma>0$ is the risk-aversion coefficient.

The objective in Eq.~\eqref{eq:objective} balances reward and stability, with $F(\Omega)$ encouraging allocations toward higher expected returns, while $R(\Omega)$ and the penalty terms discourage violating the constraint. The transaction cost term further discourages frequent reallocations. Moreover, we include transaction costs to reflect that portfolio adjustments unfold over time and that market evolution affects later decisions, rather than treating each interval as an isolated static optimization. These cross time couplings increase the problem size and place stricter requirements on coefficient precision, which motivates our focus on how the resulting binary formulation can be executed reliably on devices with limited numerical accuracy.

    In order to work with an Ising machine, we transform the objective function into a quadratic unconstrained binary optimization (QUBO) problem. A QUBO is the minimization of a quadratic polynomial over binary variables. For convenience, we convert our maximization objective into a minimization by negating it, $O_{\text{qubo}}(\Omega) = -O(\Omega)$. We express the investment weights using binary variables,
\begin{equation}
    \omega_{t,a} = \sum_{r = 0}^{N_r-1}2^r x_{t,a,r},
\end{equation}
where $x_{t,a,r}\in\{0,1\}$ and $N_r$ is the number of binary variables used to represent the investment in each asset at each time. The total number of binary variables in the resulting QUBO is $N_t\times N_a\times N_r$. The QUBO can be converted to an Ising Hamiltonian by mapping $x_{t,a,r} \mapsto (1-Z_i)/2$, where $i$ is a flattened index over $(t,a,r)$ and $Z_i$ denotes the Pauli $Z$ operator.

\subsection{Temporal block structure and separation of scales}

In the DPO formulation, the binary decision variables naturally partition by time. Each time period forms a block that contains all discretized allocation variables for that period across all assets. Under the QUBO encoding, this induces a block structured coefficient matrix in which most large magnitude interactions are confined within individual time blocks. Objective terms determined by the portfolio composition within a period, such as expected return, risk, and budget or leverage constraints, couple variables within the same period and typically generate dense intra-block interactions.

By contrast, interactions between different time blocks arise primarily from trading frictions $C(\Omega)$ that link consecutive periods, most notably transaction costs that depend on changes in positions across time. These terms couple variables at time $t$ to variables at time $t+1$, producing nonzero couplings only between adjacent time blocks. As a result, $Q$ has a block tridiagonal sparsity pattern along the temporal dimension, with off-diagonal blocks capturing the inter-period couplings. Moreover, these terms are scaled by coefficients that are substantially smaller than those associated with risk and return. As a consequence, the inter-block coefficients are often \textbf{several orders of magnitude smaller} than the intra-block coefficients.

The resulting QUBO therefore exhibits a separation of scales that can be summarized as strong intra-block interaction and weak inter-block coupling. This structure is important for both QUBO computation and implementation. It suggests that large instances can be approached by decomposing the problem along time blocks, while the comparatively small inter-block couplings may be more sensitive to coefficient scaling and finite-precision quantization.

\subsection{Finite-precision coefficient constraints
}

Portfolio rebalancing with realistic frictions can be expressed as a high dimensional, nonconvex optimization problem. Such formulations quickly lead to binary quadratic models with many variables and couplings, where obtaining optimal solutions is often impractical and one instead seeks high quality approximate solutions under time and resource constraints. This motivates interest in quantum optimization hardware that targets binary quadratic models. Ising based optimization hardware refers to computing platforms designed to minimize an Ising Hamiltonian. Since a QUBO can be mapped to an equivalent Ising form by a standard change of variables, these machines provide a natural backend for large scale combinatorial optimization in regimes where exact methods are computationally expensive.

A representative example is the coherent Ising machine (CIM), an optical oscillator based system that implements an effective Ising energy landscape and evolves toward low energy configurations. In practice, mapping a real valued QUBO or Ising formulation to an executable instance on such hardware requires respecting device level coefficient constraints. The programmable biases and couplings must lie in a bounded range and take values from a discrete set, which necessitates coefficient scaling followed by quantization. This issue is particularly pronounced for our DPO model because of its separation of scales, with strong intra-block interactions and weak inter-block coupling.

In this work, we instantiate the finite-precision constraint using the Qboson CIM backend used in our experiments. Specifically, programmable Ising coefficients are restricted to \texttt{int8} signed integers in the range $[-2^7, 2^7-1]$. Under a single global scaling, the large dynamic range can cause weak inter-block coefficients to be rounded to zero, effectively removing the temporal links between periods. Since these links encode rebalancing costs and temporal consistency of positions, their loss can change the optimization landscape and the resulting trading decisions. This motivates decomposition strategies that reduce the dynamic range within each subproblem while preserving the influence of inter-block couplings.


\section{Block coordinate descent}

\begin{figure}[htbp]
    \centering
    \includegraphics[width=0.9\linewidth]{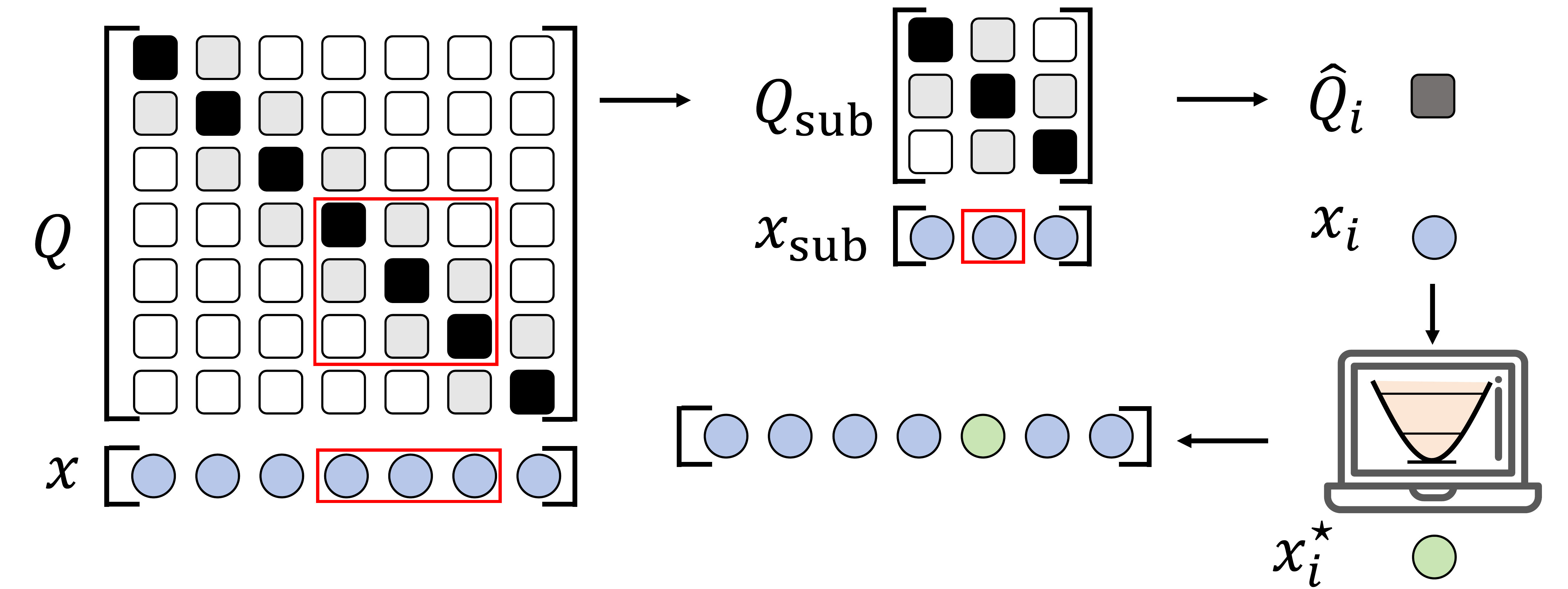}
    \caption{
    Illustration of a single BCD update. The QUBO matrix \(Q\) exhibits a block-tridiagonal structure across time steps: black diagonal blocks represent intra-block terms, gray first off-diagonal blocks represent inter-block couplings, and white blocks are zero. At block \(i\), the corresponding subproblem \((Q_{\mathrm{sub}}, x_{\mathrm{sub}})\) is extracted to construct the local QUBO \(\widehat{Q}_i\) according to Eq.~\eqref{eq:Q_hat}. The resulting subproblem is then solved for \(x_i^\star\) using either a classical solver or a CIM backend, and the solution is written back to update \(x\).}
    \label{fig:BCD scheme}
\end{figure}

Block coordinate descent (BCD) is an iterative optimization strategy that minimizes a multivariate objective by repeatedly optimizing over one block of variables while keeping the remaining variables fixed. For continuous problems, convergence properties of coordinate and block-coordinate methods have been studied extensively~\cite{tseng2001convergence,beck2013convergence}. In our discrete QUBO setting, we use BCD as a scalable heuristic. Instead of solving the full QUBO in one shot, we solve a sequence of smaller block subproblems with a classical solver or a CIM backend and insert each block solution back into the global decision vector. Figure~\ref{fig:BCD scheme} illustrates one BCD update. At each step, a local subproblem is extracted from the global QUBO, solved for the selected block while neighboring blocks are held fixed, and then written back into the global decision vector. 

\subsection{BCD general procedure}\label{subsec: BCD}

We now formalize this block update procedure for a general QUBO.
We consider a QUBO over a global binary decision vector $x \in \{0,1\}^n$ with index set $\mathcal{V}$, where $|\mathcal{V}|=n$. The index set is partitioned into $m$ non-overlapping blocks
$\mathcal{V}_1,\mathcal{V}_2,\dots,\mathcal{V}_m$ such that $\mathcal{V}=\bigcup_{k=1}^m \mathcal{V}_k,
\mathcal{V}_p \cap \mathcal{V}_q = \emptyset, \; \forall p\neq q.$ For each block $k$, we denote the corresponding subvector by $x_k = x_{\mathcal{V}_k} \in \{0,1\}^{n_k}, n_k = |\mathcal{V}_k|.$ The global QUBO objective is
\begin{equation}
    \min_{x\in\{0,1\}^n} \; f(x) = x^T Q x,
\end{equation}
where $Q\in\mathbb{R}^{n\times n}$. Without loss of generality, we assume $Q$ is symmetric. $Q$ is partitioned conformably with the blocks $\{\mathcal{V}_k\}_{k=1}^m$ as $Q_{pq} = Q_{\mathcal{V}_p,\mathcal{V}_q}, p,q\in\{1,\dots,m\}.$ In our temporal DPO instances, only adjacent time steps interact, so $Q$ is in the block tridiagonal structure $Q_{pq} = 0 \; \text{whenever } |p-q|>1.$

Choosing an index $i\in\{1,\dots,m\}$, we optimize $x_i$ while holding all other blocks fixed. Under block tridiagonal coupling, only the neighboring blocks $x_{i-1}$ and $x_{i+1}$ influence the subproblem for $x_i$. To make this explicit, define the local stacked vector and local block matrix
\begin{equation}    
x_{\text{sub}} =
\begin{bmatrix}
x_{i-1}\\ x_i\\ x_{i+1}
\end{bmatrix},
\qquad
Q_{\text{sub}} =
\begin{bmatrix}
Q_{i-1,i-1} & Q_{i-1,i}   & 0\\
Q_{i,i-1}   & Q_{i,i}     & Q_{i,i+1}\\
0           & Q_{i+1,i}   & Q_{i+1,i+1}
\end{bmatrix},
\end{equation} 
where $Q_{i,i-1}=Q_{i-1,i}^T$ and $Q_{i+1,i}=Q_{i,i+1}^T$ for symmetric $Q$. Calculating the quadratic form $x_{\text{sub}}^TQ_{\text{sub}}x_{\text{sub}}$ and collecting the terms that depend on $x_i$ yields
\begin{align}
f(x_i)
&= x_i^T Q_{ii} x_i
+ 2 x_{i-1}^T Q_{i-1,i} x_i
+ 2 x_i^T Q_{i,i+1} x_{i+1}
+ \text{const}, \label{eq:block_expand}
\end{align}
where const denotes terms independent of $x_i$ and therefore irrelevant for the minimizer of the $i$-th block update. Define the induced linear coefficient vector
\begin{equation}  
h_i = 2 Q_{i-1,i}^T x_{i-1} + 2 Q_{i,i+1} x_{i+1}.
\end{equation}
Then the $i$-th BCD subproblem can be written as the binary quadratic with linear objective
\begin{equation}   
\arg\min_{x_i\in\{0,1\}^{n_i}}
\; x_i^T Q_{ii} x_i + h_i^T x_i .
\end{equation}
Using the binary identity $x_{i,j}^2 = x_{i,j}$ for each component, the linear term can be folded into the diagonal of the quadratic form by defining 
\begin{equation}\label{eq:Q_hat}
 \widehat{Q}_i = Q_{ii} + \mathrm{diag}(h_i).   
\end{equation}
This yields an equivalent local QUBO, $x_i^\star = \arg\min_{x_i\in\{0,1\}^{n_i}} \; x_i^T \widehat{Q}_i x_i.$ Note that the boundary blocks located at the start or end of the chain follow the same update rule as interior blocks. The difference is that they have only one adjacent neighbor, so the local subproblem depends only on the block’s own quadratic terms and its coupling to that single neighboring block.

The local subproblem can be solved using either a classical solver or a CIM backend. For CIM backend, additional precision adaptation and quantization for Ising coefficients are required, which we introduce in Sec ~\ref{subsec: precision adapt}. Because many practical solvers are stochastic, we solve the same local subproblem multiple times using different random seeds. This produces multiple candidate solutions for $x_i$. For each candidate, we compute its local energy using $\widehat{Q}_i$, and we select the candidate with the lowest local energy as the accepted solution for block $i$. After obtaining $x_i^\star$ for block $\mathcal{V}_i$, we update the global decision vector by replacing only the entries in $\mathcal{V}_i$ and leaving all other entries unchanged, that is $x^{\mathrm{new}}_{\mathcal{V}_i} = x_i^\star$ and $x^{\mathrm{new}}_{\mathcal{V}\setminus \mathcal{V}_i} = x^{\mathrm{old}}_{\mathcal{V}\setminus \mathcal{V}_i}$.

The algorithm proceeds by applying the above block update sequentially for all  $i$. Completing updates for all blocks constitutes one global iteration. We repeat global iterations until reaching a prescribed maximum number of global iterations. We observe that only a small number of global iterations is typically sufficient in practice, and the resulting objective value is comparable to that obtained by applying the same solver directly to the global QUBO, when this direct solve is computationally feasible.




\subsection{Precision adaptation and quantization for hardware}\label{subsec: precision adapt}

After forming each BCD subproblem, we convert its QUBO representation to an Ising model and then adapt coefficient precision so that the resulting parameters satisfy the hardware range and finite resolution constraints. This step is necessary because direct scaling and rounding can distort small but important coefficient differences, which may change the optimizer returned by the device. Since constant offsets do not affect the minimizer, we drop additive constants and focus on the coefficients that determine the minimum.

Before quantization, we reduce the coefficient spread using the optimum-preserving scheme of \cite{mucke2025optimum}. Let $X$ denote the collection of Ising coefficients for the subproblem, including all linear and quadratic terms. Define the set of its absolute differences between all elements as $D(X) \coloneqq \{\,|a-b| : a\neq b,\ a,b\in\{X_{ij}\}\,\}$. The dynamic range~(DR) of $X$ is defined as
\begin{equation}
    DR(X) \coloneqq \log_2\left(\frac{\max D(X)}{\min D(X)} \right), 
\end{equation}
where $\max D(X)$ and $\min D(X)$ are the largest and smallest non-zero absolute differences between all elements in $X$. Large DR indicates a wide magnitude spread with very fine gaps, which is particularly sensitive to finite-precision rounding.

We apply a short sequence of single-entry tuning steps. At each step, we identify an entry that directly determines $DR(X)$, either an extreme value that contributes to $\max D(X)$ or a value involved in the tightest gap that contributes to $\min D(X)$. We then move that entry toward zero by an amount chosen from an admissible interval that simultaneously ensures DR does not increase and preserves at least one global minimizer. The update is accepted only if DR strictly decreases; otherwise the candidate is skipped. We terminate when no admissible improving move exists or after a small fixed iteration budget.

After adaptation, we quantize the coefficients to meet the integer constraint required by the hardware interface. Let $\alpha(X) \coloneqq \max_\ell |X_\ell|$ be the maximum absolute coefficient magnitude. We scale to the signed 8-bit range by applying $\widetilde{X} \;=\; \mathrm{clip}\!\left(\mathrm{round}\!\left(\frac{127}{\alpha(X)}\,X\right),\, -2^7,\, 2^7-1\right)$, where $\mathrm{round}(\cdot)$ rounds to the nearest integer and $\mathrm{clip}(\cdot)$ saturates values outside the target range. 

This quantization step highlights a trade-off between portfolio discretization fidelity and hardware precision. Increasing the number of encoding bits \(N_r\) improves the representation of the continuous portfolio weights \(\omega_{t,a}\), but also enlarges the coefficient range of the induced Ising model through factors. Under fixed \texttt{int8} precision, this makes weaker couplings, especially inter-temporal transaction-cost terms, more susceptible to rounding. Conversely, coarser discretization is more robust to quantization but reduces allocation granularity. Our BCD strategy mitigates this trade-off by reducing the coefficient range in each hardware call.

\subsection{Complexity analysis}

For DPO, variables are partitioned along the time dimension, so $m = N_t$ and each block corresponds to one time step. With $N_a$ assets and $N_r$ binary encoding bits per asset, the global number of binary variables is $n = N_a N_r N_t$. Since QUBO is NP-hard, the worst-case runtime of QUBO algorithms can scale exponentially in the number of optimized variables. If the global QUBO is solved directly, the worst-case scaling is $O\!\left(2^{N_a N_r N_t}\right).$ For BCD, each subproblem has dimension $n_i = N_a N_r$, and each global iteration solves $N_t$ subproblems of size $N_a N_r$. Denoting by $I$ the number of repeated solver runs per block due to random seeds and by $J$ the maximum number of global iterations, the corresponding worst-case scaling is $O\!\left(JN_TI\,2^{N_a N_r}\right).$ Thus, BCD replaces one global exponential search variables with a sequence of smaller subproblems of size, repeated over time blocks and outer iterations.

\section{Evaluation}
\subsection{Experiment setup}
We evaluate the proposed QUBO formulation and BCD workflow on dynamic portfolio optimization instances constructed from market data spanning 22 trading intervals and 6 assets. Details for the data are in Appendix~\ref{appendix: data prepare}.  Table~\ref{tab:custom_dpo_sizes} summarizes the configurations used throughout this section, including the number of time steps $N_t$, number of assets $N_a$, discretization level $N_r$, resulting QUBO dimension $N_q$, and the budget parameter $K$.

\begin{table}[H]
\centering
\begin{tabular}{|c|c|c|c|c|c|}
\hline
\textbf{Size} & \textbf{$N_t$} & \textbf{$N_a$} & \textbf{$N_r$} & \textbf{$N_q$} & \textbf{K} \\
\hline
S  & 2     & 6     & 4     & 48    & 15                 \\
\hline
M & 6     & 6     & 4     & 144   & 15                 \\
\hline

L  & 22    & 6     & 4     & 528   & 15                 \\
\hline

\end{tabular}
\caption{
Problem configurations used in the experiments, reporting the number of time steps \(N_t\), assets \(N_a\), binary variables per asset and time step \(N_r\), total number of binary variables \(N_q\), and budget \(K\).}
\label{tab:custom_dpo_sizes}
\end{table}

CIMs implement networks of degenerate optical parametric oscillators to represent Ising spin states, with programmable spin--spin couplings realized through an FPGA-based measurement-feedback loop. In our experiments, QUBO instances are mapped to the corresponding Ising form and solved on a 550-spin CIM platform. To contextualize solution quality, we compare CIM with three classical baselines. For small QUBO subproblems, we compute exact reference solutions using the SCIP Optimization Suite~\cite{hojny2025scip}. For all problem sizes, we evaluate simulated annealing (SA) and tabu search (TS), implemented through the Kaiwu SDK~\cite{qboson_kaiwu_sdk_2025}.

To isolate the impact of block decomposition and coefficient quantization, we report results under four variants defined by two binary choices. The first choice is whether we use the BCD block decomposition to solve a sequence of block subproblems~(denoted as Block) or solve a single global problem directly~(denoted as Global). The second choice is whether the Ising coefficients after mapping are kept in full precision~(denoted as FP) or quantized to \texttt{int8}~(denoted as INT8) using the approach in Sec.~\ref{subsec: precision adapt}. We denote the four variants as Global--FP, Block--INT8, Global--INT8, and Block--FP. For experiments on CIM, coefficients are always quantized to \texttt{int8} due to hardware constraints, so CIM is evaluated only on the INT8 variants.

Solution evaluation is based on both feasibility and performance. Feasibility is determined by the budget constraint in Eq.~\eqref{eq:budget-constraints}: solutions that violate this constraint are treated as infeasible and are discarded from performance reporting. To assess performance over the rebalancing horizon, we report the net mean return at each rebalancing time step. Let $F_t(\Omega)$ and $C_t(\Omega)$ denote the mean return and transaction cost evaluated at time step $t$, which are calculated by 
\begin{equation}
     F_t(\Omega)=\sum_{a=0}^{N_a-1}\omega_{t,a}\,\mu_{t,a},\quad
     C_t(\Omega)= \nu\lambda \sum_{a=0}^{N_a-1}\left(\omega_{t,a}-\omega_{t-1,a}\right)^2.
\end{equation}
We define the net mean return per time step as \begin{equation}
    F^{\mathrm{net}}_t(\Omega) = F_t(\Omega) - C_t(\Omega).
\end{equation} 
This quantity measures the portfolio return after accounting for transaction costs at rebalancing time $t$. We compare solvers and strategies by examining the resulting time series $\{\mathcal{F}_t(\Omega)\}_{t=0}^{N_t-1}$, where consistently higher values indicate better performance over the rebalancing horizon.
We additionally report the Sharpe ratio commonly evaluated by~\cite{sharpe1966mutual}, \begin{equation}
    S(\Omega) = \frac{F(\Omega)}{\sqrt{R(\Omega)}}.
\end{equation}
When comparing Sharpe ratios across solvers, we keep the covariance construction fixed and only compare results produced under the same risk model.

\subsection{Results}

We begin by comparing solution quality and runtime across solvers and the four solving strategies described above. We report results under the covariance risk model in the main text, using it as a representative setting for validating the proposed algorithmic framework. Additional results for the semicovariance and shrinkage covariance risk models are provided in Appendix~\ref{appendix:more results} and exhibit the same qualitative trends.

Table~\ref{tab:tab_summary_covariance} reports the runtime and Sharpe ratio for the instances considered in this study. 
\begin{table}[h]
\centering
\scalebox{0.9}{
\begin{tabular}{cccccccc}
\toprule
\multirow{2}{*}{Solver} & \multirow{2}{*}{Strategy} & \multicolumn{2}{c}{Matrix dim  48} & \multicolumn{2}{c}{Matrix dim  144} & \multicolumn{2}{c}{Matrix dim  528} \\
\cmidrule(lr){3-4}\cmidrule(lr){5-6}\cmidrule(lr){7-8}
 &  & Runtime (s) & Sharpe ratio & Runtime (s) & Sharpe ratio & Runtime (s) & Sharpe ratio \\
\midrule
\texttt{SCIP} & Block--FP & 196 & 5.10 & 667 & 8.68 & 2492 & 15.34 \\
\midrule
\texttt{TS}   & Global--FP  & 5 & 4.77 & 18 & 8.23 & 290 & 14.14 \\
\texttt{TS}   & Block--INT8 & 19 & 4.74 & 63  & 6.18 & 267 & 10.73 \\
\texttt{TS}   & Block--FP   & 15 & 4.97 & 111 & 8.54 & 295 & 15.35\\
\midrule
\texttt{SA}   & Global--FP  & 608 & 5.09 & 1784 & 8.04 & 2210 & 14.50 \\
\texttt{SA}   & Block--INT8 & 134 & 4.59 & 411   & 7.21 & 1346 & 9.14 \\
\texttt{SA}   & Block--FP  & 113 & 5.09  & 421  & 8.69 & 1341 & 15.25 \\
\midrule
\texttt{CIM}  & Block--INT8 & $\bm{2 e^{-6}}$ & 4.83 & $\bm{4e^{-6}}$ & 6.40 & $\bm{2 e^{-5}}$ & 4.56 \\
  & Global--INT8 & \multicolumn{2}{c}{Infeasible} & \multicolumn{2}{c}{Infeasible} &\multicolumn{2}{c}{Infeasible} \\
\bottomrule
\end{tabular}}
\caption{
Comparison of runtime and Sharpe ratio under the covariance risk model. The Global--INT8 variants of TS, SA, and CIM were infeasible, meaning that the resulting solutions violated the budget constraint.
}
\label{tab:tab_summary_covariance}
\end{table}
For Global--INT8, all solvers produce infeasible solutions that violate the budget constraint in Eq.~\eqref{eq:budget-constraints}, indicating that coefficient quantization at the global scale can distort the penalty landscape and hinder constraint satisfaction. In contrast, applying BCD yields feasible solutions for Block--INT8 across all solvers, which suggests that our BCD approach improves numerical conditioning under quantization and makes the budget constraint easier to enforce. For the classical solver SA, BCD also reduces runtime, in line with our complexity analysis favoring lower dimensional block updates. Noticeably, CIM is markedly faster than the classical baselines by several orders of magnitude on the INT8 settings. This gap highlights the potential of specialized quantum and quantum-inspired devices to deliver substantial speedups on large scale combinatorial optimization when combined with hardware aligned formulations and solution workflows.

\begin{figure}[h]
    \centering
    \includegraphics[width=0.6\linewidth]{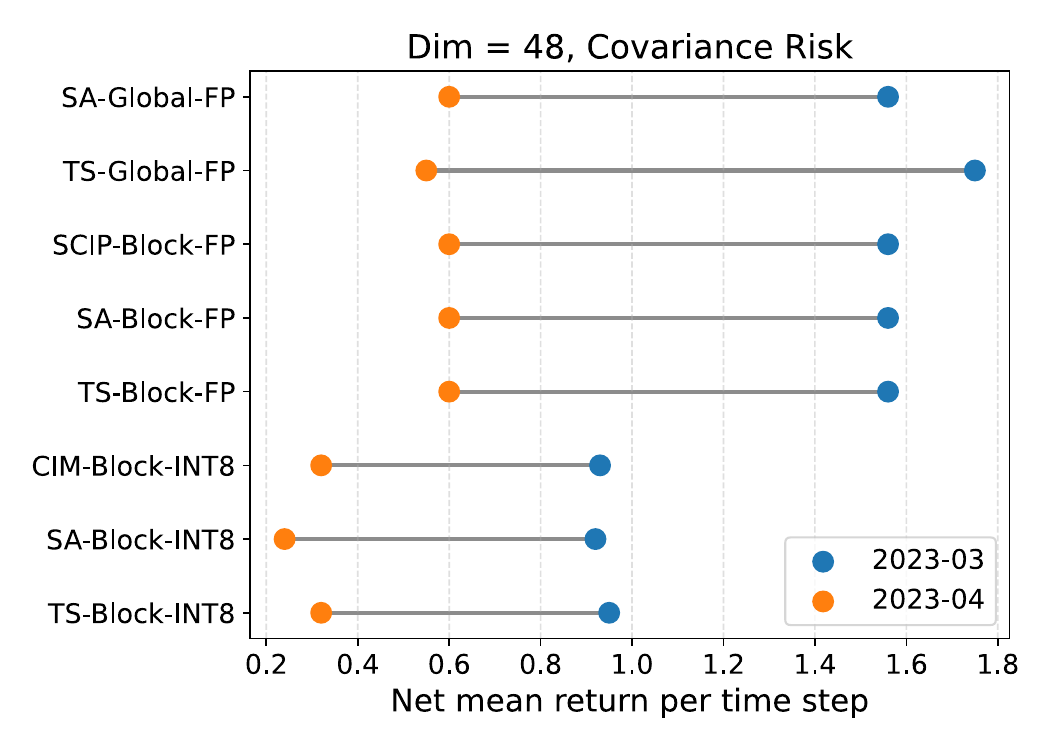}
    \caption{
    Comparison of net mean return under the covariance risk model for the dimension-48 DPO instance.
    }
    \label{fig:dim48-cov1}
\end{figure}

\begin{figure}[h]
    \centering
    \includegraphics[width=1.05\linewidth]{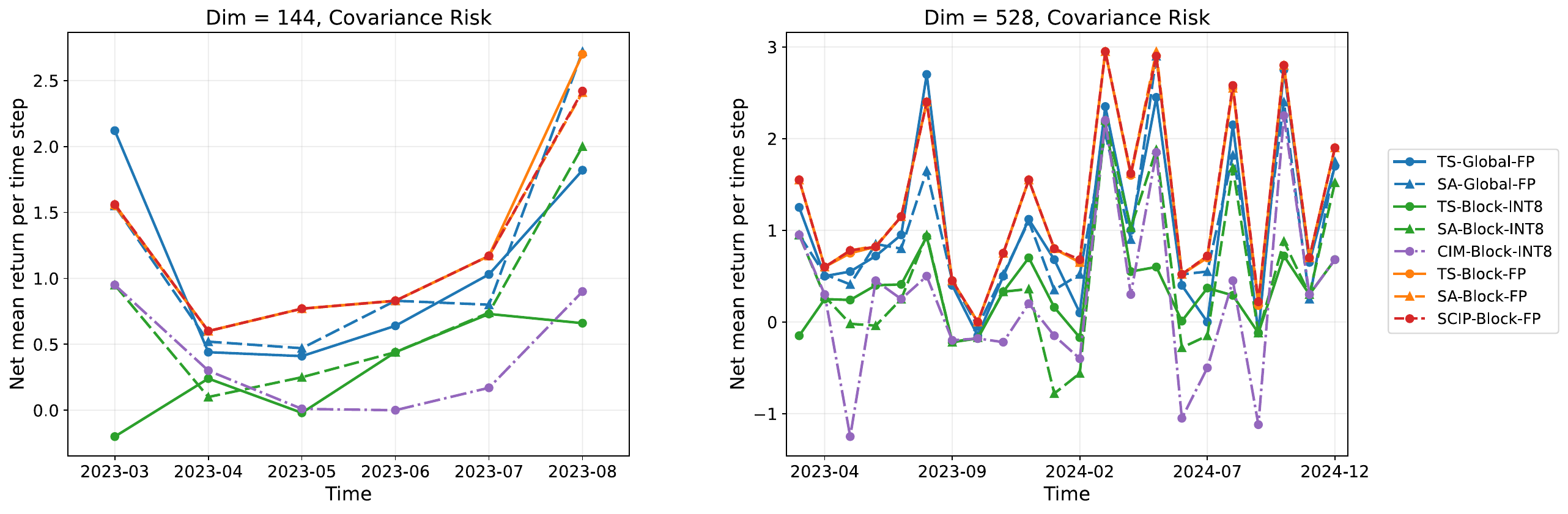}
    \caption{Comparison of net mean return over time under the covariance risk model for the dimension-144 and dimension-528 DPO instances.}
    \label{fig:dim144528-cov1}
\end{figure}

Fig.~\ref{fig:dim48-cov1} and~\ref{fig:dim144528-cov1} compare the net mean return over time for different solvers and strategy combinations. In most rebalancing steps, the full-precision variants achieve higher net mean returns than the INT8 variants, which is consistent with quantization perturbing the effective objective landscape and limiting the attainable solution quality under the same runtime budget. Comparing Global--FP and Block--FP, the return profiles are broadly similar, and neither strategy shows a systematic advantage across all times and problem sizes. Since both strategies optimize the same full-precision objective, the observed differences mainly reflect the behavior of the underlying approximate solvers under finite iteration budgets and parameter settings, rather than a change in the optimization target. When the global solve reaches a good solution within the allotted iterations, BCD typically achieves comparable returns and may in some cases be more stable under limited iterations, because each update is restricted to a lower-dimensional subproblem. Overall, the net mean return differences between Global--FP and Block--FP are modest. This observation is also consistent with the Sharpe ratios in Table~\ref{tab:tab_summary_covariance}, which indicate comparable risk-adjusted performance among the full-precision strategies.

In this setting, the reported speedups should be interpreted primarily as hardware-level reductions in solver time once a subproblem can be executed on the CIM. The main contribution of our BCD workflow is therefore not a systematic gain in return from blockwise optimization itself. Its value lies in making large DPO instances compatible with finite-precision CIM hardware by reducing both subproblem size and coefficient range, while maintaining solution quality close to the full-precision baselines. The block strategy is especially useful in the INT8 regime, where it helps enable feasible optimization and allows quantum hardware to serve as an efficient solver for larger-scale instances.

For the problem sizes considered here, the time-block subproblems produced by BCD remain within the effective capacity of the 550-spin CIM in a single hardware call. As the number of assets increases, however, the admissible block size is ultimately constrained by the available hardware capacity per call. If a time-block subproblem exceeds the capacity of a single hardware call, then an additional decomposition along the asset dimension would be required, naturally leading to a spatiotemporal extension of BCD.

Appendix~\ref{appendix:more results} shows that the same qualitative conclusions hold for the semicovariance and shrinkage covariance risk models. Although the degree of degradation under INT8 quantization varies somewhat across risk models, the overall qualitative pattern remains stable. Quantization lowers solution quality relative to full precision, while BCD consistently improves feasibility and robustness for CIM-based optimization.

\section{Conclusion and outlook}

We consider dynamic portfolio optimization (DPO) over a rebalancing time horizon and reformulate it as a QUBO and the corresponding Ising model, enabling execution on coherent Ising machines (CIMs). Under the coefficient constraints imposed by CIMs, naive global scaling and quantization can obliterate weak temporal couplings and severely degrade solution quality. To address this precision bottleneck, we propose a block coordinate descent (BCD) method tailored to the hardware limitations. The method decomposes the global QUBO along the time dimension, repeatedly solves compact per-block subproblems on the device, and accounts for neighboring time blocks by treating their current assignments as fixed during each subproblem solve. This reduces both the effective QUBO problem size and the dynamic range that each device call must represent, thereby mitigating the need for aggressive global scaling that would otherwise destroy inter-block information. Combined with coefficient adaptation and a hardware-compatible quantization scheme, the approach provides a practical way to preserve temporal coupling effects under finite-precision constraints while leveraging the runtime speedups offered by quantum hardware.

Future work should further integrate precision awareness into the full solution pipeline. Promising directions include joint optimization over short time windows to better capture inter-period structure, as well as adaptive block schedules and warm starts to improve convergence. It is also of interest to transfer the same decomposition principle to gate-based quantum solvers by using BCD to reduce instance size per call and improve accuracy in variational quantum algorithms such as the variational quantum eigensolver (VQE) and the quantum approximate optimization algorithm (QAOA).

\section{Acknowledgements}
This work was supported by the National Key R\&D Program of China (Grant No.~2024YFB4504004), the CCF-QBoson Quantum Computing Application Innovation Fund (Grant. No.~CCF-QBoson202405), the National Natural Science Foundation of China (Grant. No.~92576114, 12447107, 62302346), and the Guangdong Provincial Quantum Science Strategic Initiative (Grant No.~GDZX2403008, GDZX2503001). 


\bibliographystyle{alpha}
\bibliography{ref}


\newpage

\appendix
\setcounter{subsection}{0}
\setcounter{table}{0}
\setcounter{figure}{0}

\vspace{3cm}

\begin{center}
\Large{\textbf{Supplemental Material --- Block Coordinate Descent for Dynamic Portfolio Optimization on Finite-Precision Coherent Ising Machines}}
\end{center}

\renewcommand{\theequation}{S\arabic{equation}}
\renewcommand{\theproposition}{S\arabic{proposition}}
\renewcommand{\thedefinition}{S\arabic{definition}}
\renewcommand{\thefigure}{S\arabic{figure}}


\section{Data description and preprocessing}
\label{appendix: data prepare}

\begin{figure}[h]
    \centering
    \includegraphics[width=0.9\linewidth]{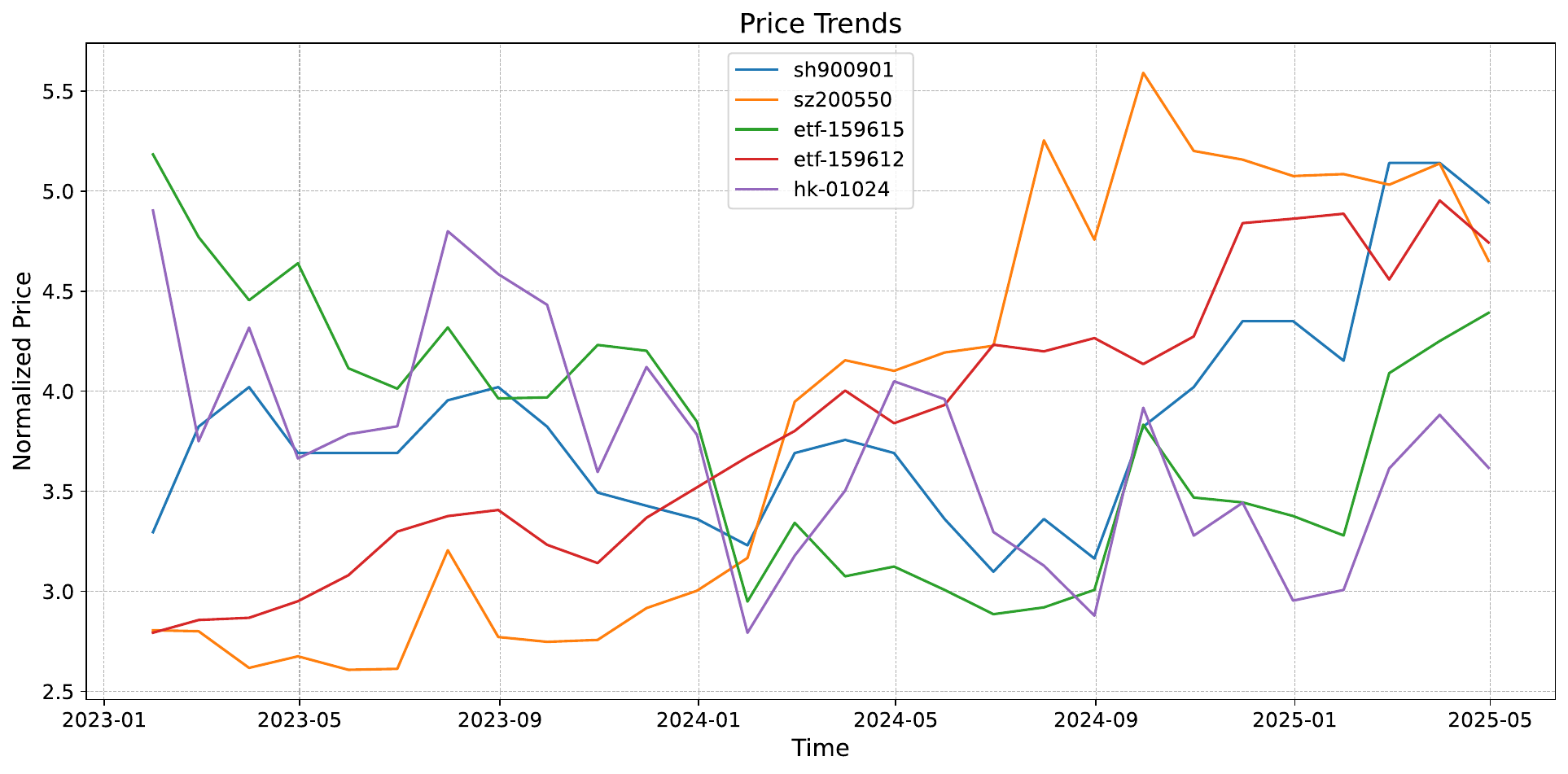}
    \caption{Normalized closing-price trajectories of five selected assets.}
    \label{fig:price_trends}
\end{figure}

The experiments use closing price data for six assets from January 1, 2023 to May 1, 2025, yielding 515 valid trading days and 22 trading intervals after preprocessing. The asset set is chosen as a realistic and manageable testbed for validating the proposed algorithmic framework. Most assets are selected from the financial markets of mainland China and Hong Kong, with one additional no-trade asset representing cash and modeled by a constant price series. Since the assets have different price scales, we normalize each closing price series and focus on relative price changes rather than absolute price levels. Fig.~\ref{fig:price_trends} shows the normalized price trajectories of the assets.

\section{Solver settings, hardware platform, and evaluation protocol}\label{appendix:more results}

We summarize here the solver settings and execution protocol used in the numerical experiments. For classical solvers, we use the mixed-integer quadratic programming~(MIQP) solver in the SCIP Optimization Suite~\cite{hojny2025scip}, as well as simulated annealing~(SA) and tabu search~(TS) implemented through the Kaiwu SDK~\cite{qboson_kaiwu_sdk_2025}. All classical experiments are executed on a server equipped with an AMD EPYC 9554 64-Core CPU. The CIM results are obtained using the CPQC-550 coherent Ising machine~(CIM) platform provided by Beijing QBoson Quantum Technology Co., Ltd.

To mitigate stochastic variability, each BCD update solves the same block subproblem 3 times and retains the solution with the lowest local energy. We run BCD for 3 global iterations. For each Block and Global strategy, we perform 3 independent runs and report the run with the highest net mean return.

Additional results for the semicovariance and shrinkage covariance risk models are reported in Table~\ref{tab:tab_summary_Semicovariance} and Table~\ref{tab:tab_summary_Shrinkagecovariance}, respectively.

\begin{table}[h]
\centering
\scalebox{0.9}{
\begin{tabular}{cccccccc}
\toprule
\multirow{2}{*}{Solver} & \multirow{2}{*}{Strategy} & \multicolumn{2}{c}{Matrix dim  48} & \multicolumn{2}{c}{Matrix dim  144} & \multicolumn{2}{c}{Matrix dim  528} \\
\cmidrule(lr){3-4}\cmidrule(lr){5-6}\cmidrule(lr){7-8}
 &  & Runtime (s) & Sharpe ratio & Runtime (s) & Sharpe ratio & Runtime (s) & Sharpe ratio \\
\midrule
\texttt{SCIP} & Block--FP & 223 & 4.99 & 629 & 8.41 & 2368 & 15.43 \\
\midrule
\texttt{TS}   & Global--FP  & 7 & 5.01 & 17 & 8.02 & 282 & 14.34 \\
\texttt{TS}   & Block--INT8 & 25 & 4.63 & 68  & 6.77 & 271 & 10.29 \\
\texttt{TS}   & Block--FP   & 14 & 5.02 & 113 & 8.50 & 287 & 15.35\\
\midrule
\texttt{SA}   & Global--FP  & 506 & 5.05 & 1313 & 8.28 & 1959 & 13.60 \\
\texttt{SA}   & Block--INT8 & 138 & 3.35 & 410   & 7.70 & 1346 & 11.10 \\
\texttt{SA}   & Block--FP  & 117 & 5.05  & 431  & 8.37 & 1335 & 15.45\\
\midrule
\texttt{CIM}  & Block--INT8 & $\bm{2 e^{-6}}$ & 4.80 & $\bm{4 e^{-6}}$ & 3.04 & $\bm{2 e^{-5}}$ & 2.50 \\
  & Global--INT8 & \multicolumn{2}{c}{Infeasible} & \multicolumn{2}{c}{Infeasible} &\multicolumn{2}{c}{Infeasible} \\
\bottomrule
\end{tabular}}
\caption{Comparison of runtime and Sharpe ratio under the semicovariance risk model. The Global--INT8 variants of TS, SA, and CIM were infeasible, meaning that the resulting solutions violated the budget constraint. }
\label{tab:tab_summary_Semicovariance}
\end{table}

\begin{table}[h]
\centering
\scalebox{0.9}{
\begin{tabular}{cccccccc}
\toprule
\multirow{2}{*}{Solver} & \multirow{2}{*}{Strategy} & \multicolumn{2}{c}{Matrix dim  48} & \multicolumn{2}{c}{Matrix dim  144} & \multicolumn{2}{c}{Matrix dim  528} \\
\cmidrule(lr){3-4}\cmidrule(lr){5-6}\cmidrule(lr){7-8}
 &  & Runtime (s) & Sharpe ratio & Runtime (s) & Sharpe ratio & Runtime (s) & Sharpe ratio \\
\midrule
\texttt{SCIP} & Block--FP & 250 & 5.06 & 647 & 8.66 & 2492 & 15.30 \\
\midrule
\texttt{TS}   & Global--FP  & 6 & 4.96 & 10 & 8.72 & 290 & 13.42 \\
\texttt{TS}   & Block--INT8 & 21 & 4.49 & 58  & 7.83 & 271 & 9.87 \\
\texttt{TS}   & Block--FP   & 14 & 5.13 & 112 & 8.54 & 291 & 15.33\\
\midrule
\texttt{SA}   & Global--FP  & 702 & 5.14 & 1222 & 8.39 & 2349 & 15.03 \\
\texttt{SA}   & Block--INT8 & 133 & 4.78 & 410   & 8.53 & 1346 & 10.72 \\
\texttt{SA}   & Block--FP  & 122 & 5.14  & 412  & 8.63 & 1341 & 15.36 \\
\midrule
\texttt{CIM}  & Block--INT8 & $\bm{2 e^{-6}}$ & 3.33 & $\bm{4 e^{-6}}$ & 0.17 & $\bm{2 e^{-5}}$ & 2.93 \\
  & Global--INT8 & \multicolumn{2}{c}{Infeasible} & \multicolumn{2}{c}{Infeasible} &\multicolumn{2}{c}{Infeasible} \\
\bottomrule
\end{tabular}}
\caption{Comparison of runtime and Sharpe ratio under the shrinkage covariance risk model. The Global--INT8 variants of TS, SA, and CIM were infeasible, meaning that the resulting solutions violated the budget constraint. }
\label{tab:tab_summary_Shrinkagecovariance}
\end{table}

\begin{figure}[h]
    \centering
    \includegraphics[width=0.6\linewidth]{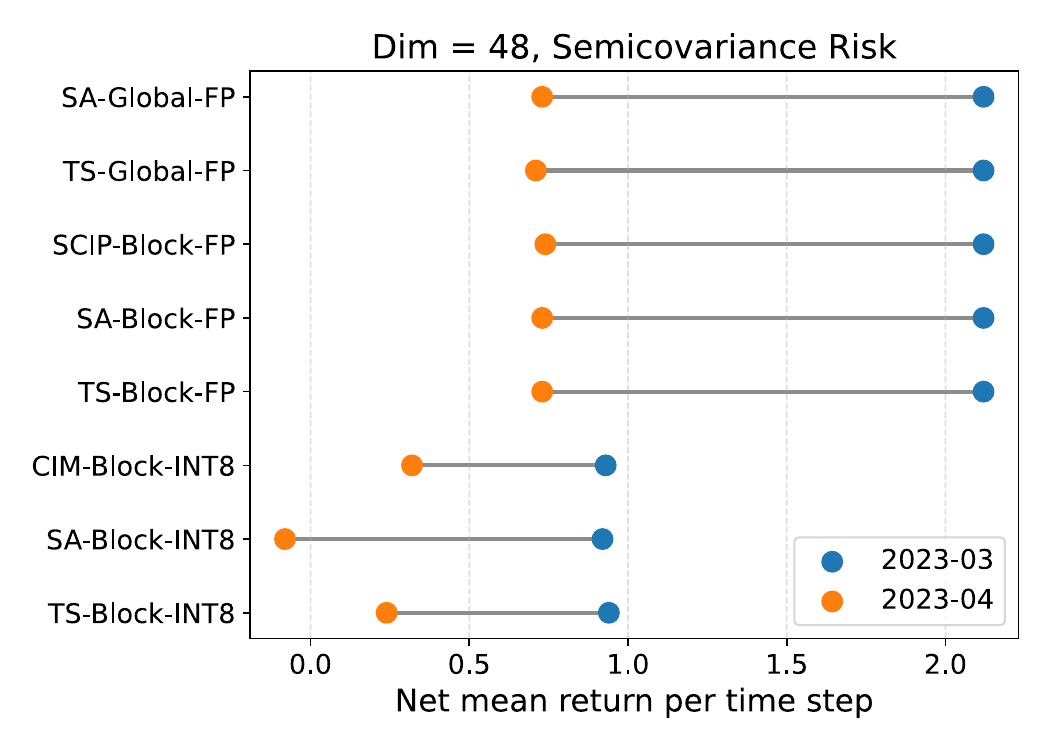}
    \caption{Comparison of net mean return under the semicovariance risk model for the dimension-48 DPO instance.}
    \label{fig:dim48-cov2}
\end{figure}

\begin{figure}[h]
    \centering
    \includegraphics[width=1.05\linewidth]{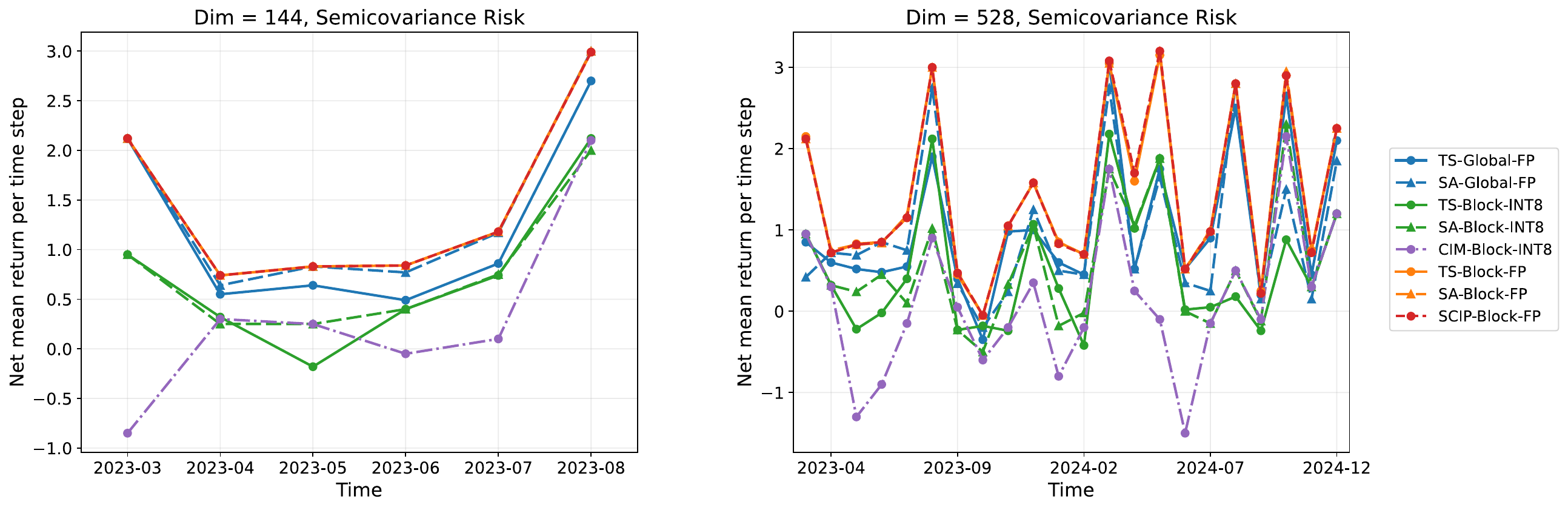}
    \caption{Comparison of net mean return over time under the semicovariance risk model for the dimension-144 and dimension-528 DPO instances. }
    \label{fig:dim144528-cov2}
\end{figure}

\begin{figure}[h]
    \centering
    \includegraphics[width=0.6\linewidth]{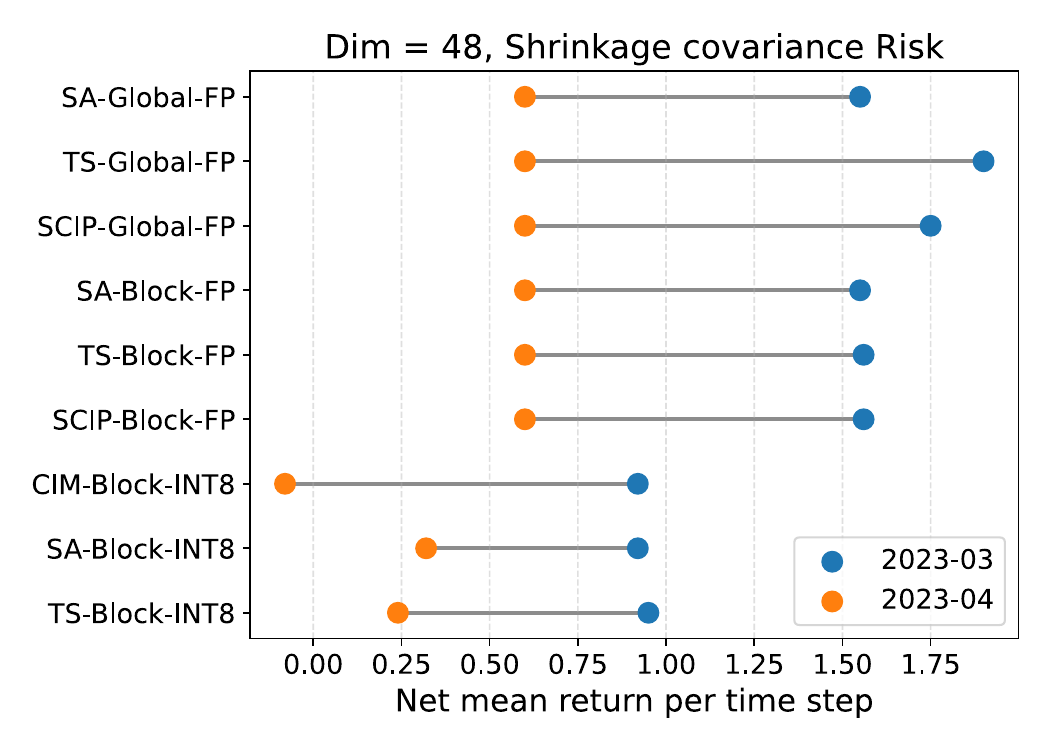}
    \caption{Comparison of net mean return under the shrinkage covariance risk model for the dimension-48 DPO instance.}
    \label{fig:dim48-cov3}
\end{figure}

\begin{figure}[h]
    \centering
    \includegraphics[width=1.05\linewidth]{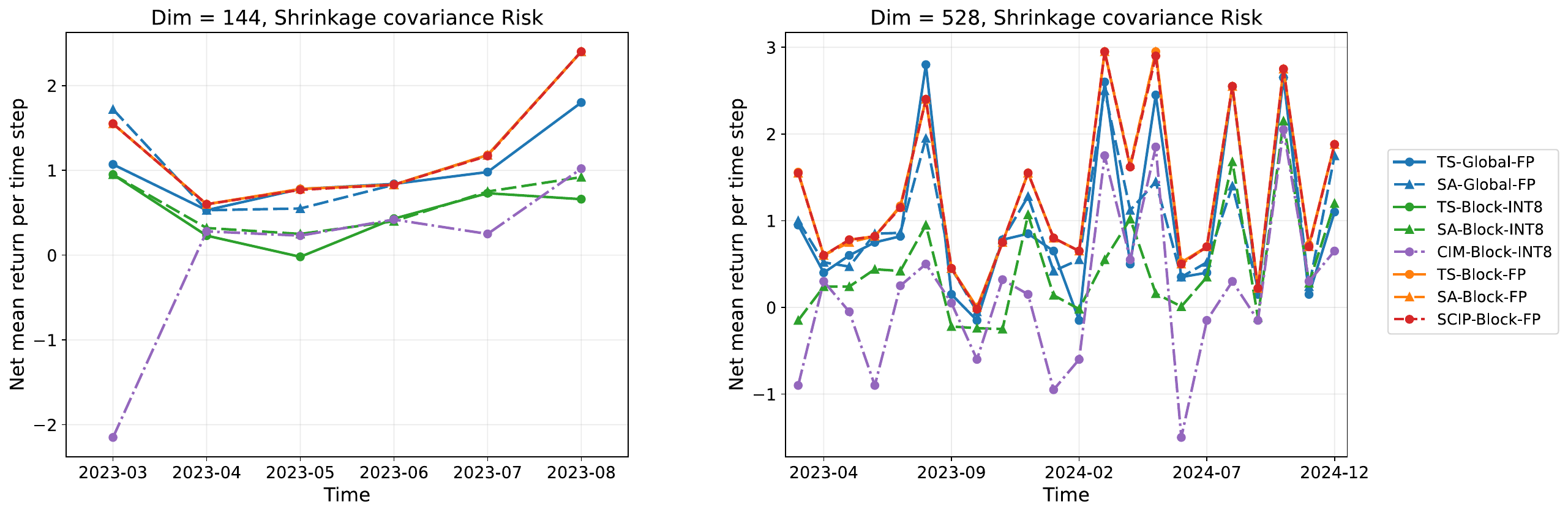}
    \caption{Comparison of net mean return over time under the shrinkage covariance risk model for the dimension-144 and dimension-528 DPO instances. }
    \label{fig:dim144528-cov3}
\end{figure}

Fig.~\ref{fig:dim48-cov2} and~\ref{fig:dim144528-cov2} compare the net mean return over time for different solvers and strategy combinations under the semicovariance risk model. Fig.~\ref{fig:dim48-cov3} and~\ref{fig:dim144528-cov3} report the corresponding results under the shrinkage covariance risk model.

\end{document}